# Semantic Numeration Systems as Dynamical Systems

**Alexander Yu. Chunikhin**
Palladin Institute of Biochemistry
National Academy of Sciences of Ukraine
alexchunikhin61@gmail.com
ORCID 0000-0001-8935-0338

**Abstract**. The foundational concepts of semantic numeration systems theory are briefly outlined. The action of cardinal semantic operators unfolds over a set of cardinal abstract entities belonging to the cardinal semantic multeity. The cardinal abstract object (CAO) formed by them in a certain connectivity topology is proposed to be considered as a linear discrete dynamical system with nonlinear control. Under the assumption of ideal observability, the CAO state equations are provided for both stationary and non-stationary cases. The fundamental role of the configuration matrix, which combines information about the types of cardinal semantic operators in the CAO, their parameters and topology of connectivity, is demonstrated.

### 1. Introduction

The importance of numeration systems as fundamental pillars in the implementation of computational processes is difficult to overestimate. Currently, there are a significant number of different numeration systems [Rigo, 2014], the majority of which are predominantly theoretical. There are relatively few works that consider numeration systems from the perspective of dynamical systems theory. Without claiming completeness, we would like to cite the most significant works in our opinion.

In the series of papers, see, for example, the fourth part [Akiyama et al., 2008], the authors investigate families of dynamical systems which are related to generalized radix representations. The properties of these dynamical systems lead to new results on the characterization of bases of Pisot Number Systems as well as Canonical Number Systems. The main part is devoted to the systematic study of the so-called shift radix systems. [Barat et al., 2006] offer a consistent presentation of numeration from a dynamical viewpoint: numeration systems, their associated compactification, and dynamical systems that can be naturally defined on them. The survey [Berthé, 2012] aims to provide both a dynamical and computer arithmetic-oriented presentation of several classical numeration systems, focusing on the discrete dynamical systems that underlie them. In [Kamae, 2006] a numeration system is shown to imply a compactification of real numbers as a result of the digitalization.

In all the above works, the representation of numbers in a certain numeration system is treated as having been accomplished, completed. In the proposed work, the representation of a number (more precisely, a multinumber) is presented as a sequence of transformations – a dynamic process in a certain state space.

It should also be noted that the works mentioned above consider only the linear topology of numeration systems. This implies a condition (often left implicit) that the value of a higher (or subsequent) digit in a number representation directly depends on the value of a lower (or preceding) digit. In the semantic numeration systems considered below, the value of a subsequent digit is generally determined by more than one preceding digit with different radices.

## 2. The theoretical foundation of SNS

This section presents the basic concepts of the semantic numeration systems theory [Chunikhin, 2021, 2022].

An *abstract entity* (Æ) is an entity of arbitrary nature provided with an identifier name that allows it to be distinguished from other entities. *Cardinal Abstract Entity* (CÆ) is an abstract entity with a cardinal characteristic $CÆ_i = (i; \#_i)$, where $i$ is the name of the cardinal abstract entity, $\#_i = \text{Card}(CÆ_i)$, $\#_i \in \mathbf{N_0}$.

*Multeity* is the manifestation of something essentially uniform in various kinds and forms as well as the quality or condition of being multiple or consisting of many parts. Since we will further deal with the transformation of meanings, we define the corresponding specific type of multeity as semantic. *Semantic multeity* is an abstract space with no more than a countable set of abstract entities, semantically united by a context.

*Cardinal Semantic Multeity* (CSM) is a semantic multeity, each element of which is equipped with a cardinal characteristic - the multiplicity of a given abstract entity represented in multeity. From a set-theoretic point of view, a cardinal semantic multeity is a multiset, the carrier of which is contextually conditioned. The elements of the cardinal semantic multeity are cardinal abstract entities.

*Cardinal Semantic Operator* (CSO) is a multivalued mapping of the cardinal semantic multeity on itself, which associates a set of entity-operands from the multeity with a set of entity-images from the same multeity, transforming their cardinals using the operations defined by the operator signature: $\text{Signt}(CSO) = (K, Form, |n\rangle_w, |r\rangle_v)$, where $K$ is the operator kind, *Form* is the operator type, $|n\rangle$ is a radix-vector, $|r\rangle$ is a conversion vector. The pair (W, V) is a valence of the cardinal semantic operator.

The main *forms* of the cardinal semantic operators are:

➢ *L-operator* (Line-operator): ($\uparrow\#$, L, $n_i$, $r_{ij}$) – a cardinal semantic operator of valency (W, V) = (1, 1), which, to each unit of carry $p_i$ (radix-ni multiplicity) from the cardinal abstract entity $CÆ_i$, assigns (gives the meaning of) $r_{ij}$ units of the transformant $q_j$ added to the cardinal $\#_j$ of the abstract entity $CÆ_j$.

A diagrammatic representation of the L-operator is shown in Fig. 1.

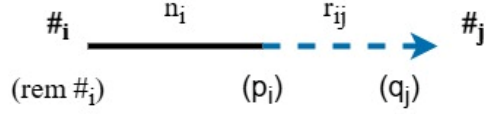

Fig. 1

When the L-operator acts on the CÆ$_i$-operand, the following operations are performed:

(i) $p_i = \lfloor \#_i/n_i \rfloor$ – calculation of the radix-multiplicity, that is, i-carry value. $\lfloor . \rfloor$ is the floor function;

(ii) $\#_i` = \text{rem } \#_i = \#_i - p_i \cdot n_i = \#_i \bmod n_i$ – finding the remainder in CÆ$_i$;

(iii) $q_j = p_i \cdot r_{ij}$ – calculation of the j-transformant value;

(iv) $\#_j` = \#_j + q_j$ – finding the change of the CÆ$_j$-image cardinal.

➢ *D-operator* (Distribution operator): (↑#, D, $n_i$, ($r_{ij}$, …,$r_{ih}$)) is a cardinal semantic operator of valency (W, V) = (1, v), which, for each unit of carry $p_i$ from the abstract entity CÆ$_i$, assigns $v$ transformants as follows: $r_{ij}$ units of j-transformants $q_j$ for the cardinal abstract entity CÆ$_j$, …, $r_{ih}$ units of h-transformants $q_h$ for the cardinal abstract entity CÆ$_h$.

A diagrammatic representation of the D-operator D$_2$ is shown in Fig. 2.

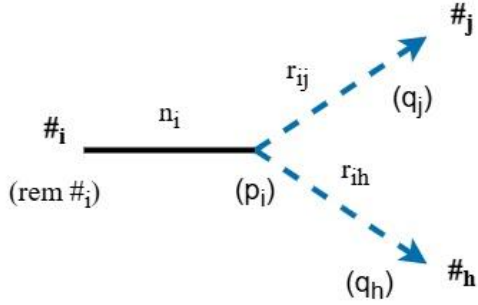

Fig. 2

➢ *F-operator* (Fusion operator): (↑#, F, ($n_i$, …,$n_j$), $r_{(ij)h}$) is a cardinal semantic operator of valency (W, V) = (w, 1), which, for each unit of *common carry* $p_{(i…j)}$ from the abstract entities CÆ$_i$, …, CÆ$_j$, assigns $r_{(ij)h}$ units of the transformant $q_h$ added to the cardinal #$_h$ of the abstract entity CÆ$_h$.

The calculation of the common carry $p_{(i…j)}$ is described below for the M-operator. A diagrammatic representation of the F-operator $_2$F is shown in Fig. 3.

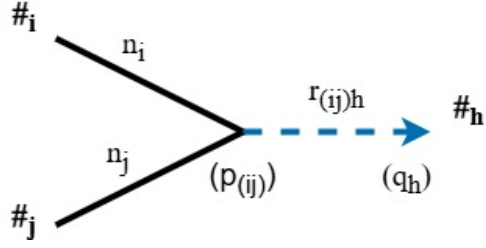

Fig. 3

➢ *M-operator* (Multi-operator): (↑#, M, ($n_i$, …,$n_j$), ($r_{(i…j)h}$, …, $r_{(i…j)g}$)) is a cardinal semantic operator of valency (W, V) = (w, v), which, for each unit of common carry $p_{(i…j)}$ from the abstract entities $CÆ_i$, …, $CÆ_j$, assigns *v* transformants as follows: $r_{(i…j)h}$ units of h-transformants $q_h$ for the cardinal abstract entity $CÆ_h$, …, $r_{(i…j)g}$ units of g-transformants $q_g$ for the cardinal abstract entity $CÆ_g$.

A diagrammatic representation of the M-operator $_2M_2$ is shown in Fig. 4.

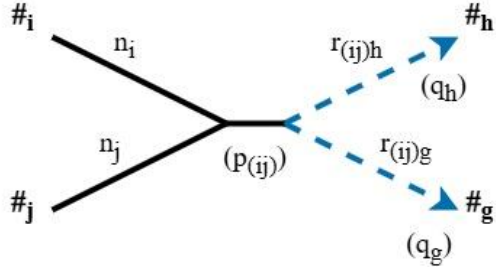

Fig. 4

When the $_2M_2$-operator acts on $CÆ_{i,j}$-operands, the following operations are performed:
(i) $p_i = \lfloor \#_i/n_i \rfloor$, $p_j = \lfloor \#_j/n_j \rfloor$ – calculation of partial carries;
(ii) $p_{(ij)} = \min\{p_i, p_j\}$ – calculation of the common carry;
(iii) $\#_i` = \#_i - p_{(ij)} \cdot n_i$, $\#_j` = \#_j - p_{(ij)} \cdot n_j$ – calculation of the remainders in $CÆ_i$ and $CÆ_j$;
(iv) $q_h = p_{(ij)} \cdot r_{(ij)h}$, $q_g = p_{(ij)} \cdot r_{(ij)g}$ – calculation of the partial transformants;
(v) $\#_h` = \#_h + q_h$, $\#_g` = \#_g + q_g$ – finding the changes of the $CÆ_h$-, $CÆ_g$-image cardinals.

These four cardinal semantic operators form the operator basis of any semantic numeration system.

To represent complex multistage semantic transformations, mono-operator transformations are typically insufficient. In [Chunikhin, 2021] the concept of a Numeration Space (NS) was introduced, the elements of which are Cardinal Abstract Objects (CAO). *A Cardinal Abstract Object* (CAO) is a set of Cardinal Abstract Entities connected in a certain topology (STop) by Cardinal Semantic Operators. A certain $CAO_I$ implements/represents a specific method of numeration *I* in a numeration space. *I* is a name (may be a complex name) denoting the $CAO_I$ as a specific numeration method in accordance with the accepted classification [Chunikhin, 2021, 2022].

*Cardinal Semantic Transformation* (CST) consists in executing all "allowed" cardinal semantic operators for the given CAO$_I$. A *CST step* means a single execution of all "allowed" cardinal semantic operators for the given CAO$_I$.

After an arbitrary step *k* of the cardinal semantic transformation, the multiset of cardinals of all CÆs from CSM is called a *multicardinal* of the CAO$_I$ of the step *k* and denoted by <A$_I$(k)> (<A$_I$(k)> = [#$_i$(k), #$_j$(k), …, #$_l$(k)] = $\langle \#(k) \rangle$). The multicardinal meaningfully characterizes only the "card-fullness" of the CAO$_I$ after each step of the cardinal semantic transformation, but in no way reflects the semantic aspect of the CST.

The holistic structural-cardinal representation of the CAO$_I$ after the *k*-th step of the cardinal semantic transformation is called the *I-multinumber* of the step *k* and denoted by A$_I$(k).

$$A_I(k) = \mathbb{L}^m(|\#(k)\rangle \mid \mathbb{E}_I),$$

where
- $\mathbb{L}^m$ is the symbol of the structural-cardinal union in the CAO$_I$ of both *m* cardinal abstract entities and the cardinal semantic operators that connect them;
- $|\#(k)\rangle$ is the *CÆ-vector* formed by *m* components of the multicardinal <A$_I$(k)>;
- $\mathbb{E}_I$ is the CAO$_I$ *configuration matrix* (*m* × *m*), containing information about the type of cardinal semantic operators in the CAO$_I$, their parameters and connectivity topology (STop).

We assume that the multicardinal determines precisely the *meaning* of the CAO after the *k*-th step of the cardinal semantic transformation, and the multinumber is its *sense*. Informally, a multinumber is a structured multicardinal, and a multicardinal is a destructured multinumber.

A *Semantic Numeration System* is defined as a class of numeration methods that are homogeneous in terms of certain classification features. For example, traditional *binary*, *decimal*, *hexadecimal* numeration systems are the corresponding numeration methods in Abstract Linear Homogeneous Natural Numeration System.

### 3. Semantic Numeration Systems as Dynamical Systems

Let us consider an arbitrary class of cardinal abstract objects as a set of structurally homogeneous dynamic systems, differing only in dimensionality and parameter values. This allows us to present the central principles of representing semantic numeration systems as dynamical systems, using an arbitrary CAO$_I$ from the given class as an example.

We define the subspace CSM$_I$ ⊆ CSM as the *m*-dimensional state space of the given CAO$_I$. Each *i*-coordinate of the space is formed by a state variable, namely, a cardinal abstract entity CÆ$_i$ ∈ CSM$_I$. The multicardinal $\langle \#(k) \rangle$ in this formulation acquires the meaning of the *state* vector of the CAOI: $|\#(k)\rangle$, $\dim(|\#(k)\rangle) = m$.

Under the assumption of ideal observability, we proceed by considering only the state equation. We will seek a solution in the class of linear discrete (digital) dynamic systems with nonlinear control [Adamy, 2024]. Discrete time $t_k = k$ is the steps of the cardinal semantic transformation (CST). A linear discrete dynamic system with nonlinear control based on a state vector is described by the following state equation:

$$x(k + 1) = A(k)x(k) + B(k)u(x(k)), \qquad (1)$$

where $x(k)$ is the *state vector*, $A(k)$ is the *state matrix*, $B(k)$ is the *control matrix*, $u(x(k))$ is the *control vector*.

### 3.1. The case of stationary CAO

At each step of the CST, the cardinal of any $CÆ_i \in CAO_I$ is determined by its current value minus what goes out as a carry, plus the transformants from the incoming CSOs. Then, for the $CAO_I$ as a whole, the state equation takes the following form:

$$|\#(k+1)\rangle = |\#(k)\rangle - N|p.(k)\rangle + R^T|p.(k)\rangle,$$

where the common carry vector $|p.(k)\rangle$ (the control vector) is defined as

$$|p.(k)\rangle = \Lambda[N^-|\#(k)\rangle].$$

Here $\Lambda$ is a *common carry operator*:

$$|p.(k)\rangle = \Lambda |p(k)\rangle.$$

The *radix operator* $N$ is a diagonal matrix ($m \times m$) with elements $n_i \delta_{ij}$. The *inverse radix operator* $N^-$ is a diagonal matrix ($m \times m$) with elements $n_i^{-1} \delta_{ij}$. The *conversion operator* $R^T$ is a matrix ($m \times m$) that contains the conversion coefficients transposed relative to their initial positions in the configuration matrix. All four of the above operators are formed based on the configuration matrix Ł.

Then, we have the following state equation of the CAO:

$$|\#(k+1)\rangle = |\#(k)\rangle + (R^T - N)\Lambda[N^-|\#(k)\rangle]. \qquad (2)$$

In cases where the CAO contains only L- and D-cardinal semantic operators, the common carry operator is not needed in the formation of the carry vector: $|p.(k)\rangle = |p(k)\rangle$.
Then,
$$|\#(k+1)\rangle = |\#(k)\rangle + (R^T - N)[N^-|\#(k)\rangle]. \qquad (3)$$

### 3.2. The case of a non-stationary CAO

The parameters of the CAO change, in general, at each step of the CST are $R^T \to R^T(k), N \to N(k), N^- \to N^-(k)$. The state equation for a non-stationary CAO takes the form:
$$|\#(k+1)\rangle = |\#(k)\rangle + (R^T(k) - N(k))\Lambda[N^-(k)|\#(k)\rangle]. \qquad (4)$$

Expressions (2), (4) are universally applicable, meaning they can be used to describe the dynamics of the CAO with an arbitrary set of cardinal semantic operators, their parameters and connectivity topology.

## 4. An Example

Let us examine in more detail the components of the state equation (2) of some $CAO_I$ using a specific example (Fig.5). We denote the initial CÆs by triangles with upward-pointing vertices, the intermediate ones by circles, and the final one by a downward-pointing triangle. This $CAO_I$ is composed of seven CÆs and four CSOs — $_2M_2$, L, $D_2$ and $_2F$ — that connect them in a given STop. For generality, all types of cardinal semantic operators are used.

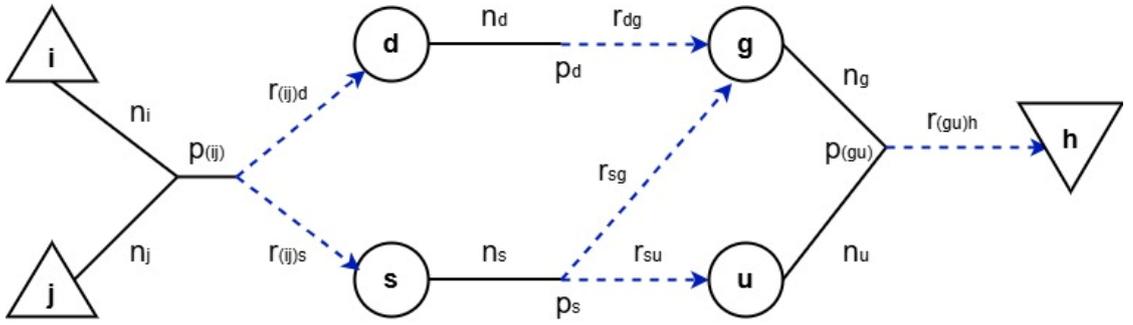

Fig. 5

The state vector (the multicardinal in vector form) for the given $CAO_I$ has the following form ($m = 7$):

$$|\#(k)\rangle = \begin{pmatrix} \#_i(k) \\ \#_j(k) \\ \#_d(k) \\ \#_s(k) \\ \#_g(k) \\ \#_u(k) \\ \#_h(k) \end{pmatrix}.$$

The configuration matrix for the considered $CAO_I$ is as follows:

$$Ł = \begin{matrix} & \backslash & i & j & d & s & g & u & h \\ i & n_i & 0 & r_{(ij)d} & r_{(ij)s} & 0 & 0 & 0 \\ j & 0 & n_j & r_{(ij)d} & r_{(ij)s} & 0 & 0 & 0 \\ d & 0 & 0 & n_d & 0 & r_{dg} & 0 & 0 \\ s & 0 & 0 & 0 & n_s & r_{sg} & r_{su} & 0 \\ g & 0 & 0 & 0 & 0 & n_g & 0 & r_{(gu)h} \\ u & 0 & 0 & 0 & 0 & 0 & n_u & r_{(gu)h} \\ h & 0 & 0 & 0 & 0 & 0 & 0 & 0 \end{matrix}.$$

From the operators $R^T$ and $N$ we immediately express the difference:

$$R^T - N = \begin{pmatrix} \backslash & i & j & d & s & g & u & h \\ i & -n_i & 0 & 0 & 0 & 0 & 0 & 0 \\ j & 0 & -n_j & 0 & 0 & 0 & 0 & 0 \\ d & r_{(ij)d} & 0 & -n_d & 0 & 0 & 0 & 0 \\ s & 0 & r_{(ij)s} & 0 & -n_s & 0 & 0 & 0 \\ g & 0 & 0 & r_{dg} & r_{sg} & -n_g & 0 & 0 \\ u & 0 & 0 & 0 & r_{su} & 0 & -n_u & 0 \\ h & 0 & 0 & 0 & 0 & r_{(gu)h} & 0 & 0 \end{pmatrix}.$$

The inverse radix operator can be written as

$$N^- = \begin{pmatrix} \backslash & i & j & d & s & g & u & h \\ i & n_i^{-1} & 0 & 0 & 0 & 0 & 0 & 0 \\ j & 0 & n_j^{-1} & 0 & 0 & 0 & 0 & 0 \\ d & 0 & 0 & n_d^{-1} & 0 & 0 & 0 & 0 \\ s & 0 & 0 & 0 & n_s^{-1} & 0 & 0 & 0 \\ g & 0 & 0 & 0 & 0 & n_g^{-1} & 0 & 0 \\ u & 0 & 0 & 0 & 0 & 0 & n_u^{-1} & 0 \\ h & 0 & 0 & 0 & 0 & 0 & 0 & 0 \end{pmatrix}.$$

The common carry operator is

$$\Lambda = \begin{pmatrix} \backslash & i & j & d & s & g & u & h \\ i & \Lambda & \Lambda & 0 & 0 & 0 & 0 & 0 \\ j & \Lambda & \Lambda & 0 & 0 & 0 & 0 & 0 \\ d & 0 & 0 & 0 & 0 & 0 & 0 & 0 \\ s & 0 & 0 & 0 & 0 & 0 & 0 & 0 \\ g & 0 & 0 & 0 & 0 & \Lambda & \Lambda & 0 \\ u & 0 & 0 & 0 & 0 & \Lambda & \Lambda & 0 \\ h & 0 & 0 & 0 & 0 & 0 & 0 & 0 \end{pmatrix}.$$

The partial carries vector $|p(k)\rangle$ is

$$|p(k)\rangle = \begin{pmatrix} p_i(k) \\ p_j(k) \\ p_d(k) \\ p_s(k) \\ p_g(k) \\ p_u(k) \\ 0 \end{pmatrix}.$$

The action of the operator that forms the common carry consists in "multiplying" the matrix $\Lambda$ by the vector of partial carries $|p(k)\rangle$ in such a way that the operation of adding the factors is replaced by the operation of taking their minimum.

Then the vector of common carries $|p.(k)\rangle$ (the control vector) is defined as (the step $k$ is omitted here):

$$\begin{pmatrix} p_{(ij)} \\ p_{(ij)} \\ p_d \\ p_s \\ p_{(gu)} \\ p_{(gu)} \\ 0 \end{pmatrix} = \begin{matrix} \\ i \\ j \\ d \\ s \\ g \\ u \\ h \end{matrix} \begin{pmatrix} i & j & d & s & g & u & h \\ \Lambda & \Lambda & 0 & 0 & 0 & 0 & 0 \\ \Lambda & \Lambda & 0 & 0 & 0 & 0 & 0 \\ 0 & 0 & 0 & 0 & 0 & 0 & 0 \\ 0 & 0 & 0 & 0 & 0 & 0 & 0 \\ 0 & 0 & 0 & 0 & \Lambda & \Lambda & 0 \\ 0 & 0 & 0 & 0 & \Lambda & \Lambda & 0 \\ 0 & 0 & 0 & 0 & 0 & 0 & 0 \end{pmatrix} \begin{pmatrix} p_i \\ p_j \\ p_d \\ p_s \\ p_g \\ p_u \\ 0 \end{pmatrix} = \begin{pmatrix} \min(p_i, p_j) \\ \min(p_i, p_j) \\ p_d \\ p_s \\ \min(p_g, p_u) \\ \min(p_g, p_u) \\ 0 \end{pmatrix}.$$

Below, we illustrate the dynamics of the given $CAO_I$ using a numerical example by parameterizing it and providing the initial data (Fig.6). Let $\#_i = 100_i, \#_j = 100_j, \forall \#_* = 0$.

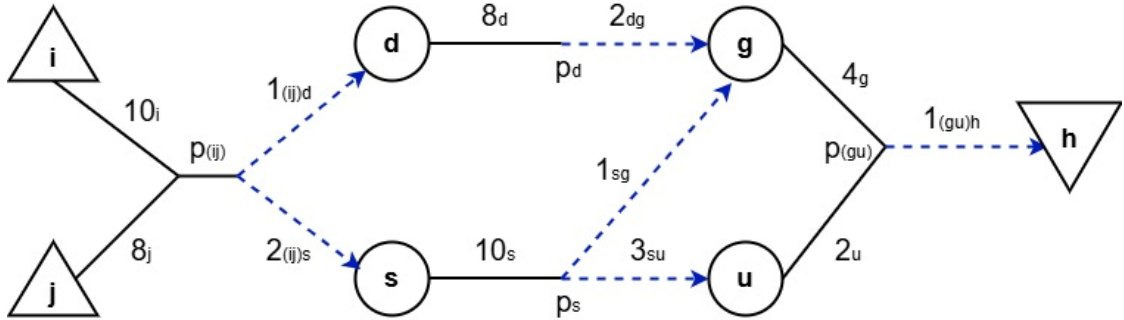

Fig. 6

This $CAO_I$ is stationary. Its configuration matrix $Ł$ is expressed as

$$Ł = \begin{matrix} \\ i \\ j \\ d \\ s \\ g \\ u \\ h \end{matrix} \begin{pmatrix} i & j & d & s & g & u & h \\ 10_i & 0 & 1_{(ij)d} & 2_{(ij)s} & 0 & 0 & 0 \\ 0 & 8_j & 1_{(ij)d} & 2_{(ij)s} & 0 & 0 & 0 \\ 0 & 0 & 8_d & 0 & 2_{dg} & 0 & 0 \\ 0 & 0 & 0 & 10_s & 1_{sg} & 3_{su} & 0 \\ 0 & 0 & 0 & 0 & 4_g & 0 & 1_{(gu)h} \\ 0 & 0 & 0 & 0 & 0 & 2_u & 1_{(gu)h} \\ 0 & 0 & 0 & 0 & 0 & 0 & 0 \end{pmatrix}.$$

Then, in accordance with the state equation (2), the state vector $|\#(k)\rangle$ at each step of the CST and the corresponding common carry vector $|p.(k)\rangle$ (control vector) are as follows:

$$
\begin{array}{cccc}
|\#(0)\rangle & |\#(1)\rangle & |\#(2)\rangle & |\#(3)\rangle \\
\begin{pmatrix} 100_i \\ 100_j \\ 0_d \\ 0_s \\ 0_g \\ 0_u \\ 0_h \end{pmatrix} \rightarrow &
\begin{pmatrix} 0_i \\ 20_j \\ 10_d \\ 20_s \\ 0_g \\ 0_u \\ 0_h \end{pmatrix} \rightarrow &
\begin{pmatrix} 0_i \\ 20_j \\ 2_d \\ 0_s \\ 4_g \\ 6_u \\ 0_h \end{pmatrix} \rightarrow &
\begin{pmatrix} 0_i \\ 20_j \\ 2_d \\ 0_s \\ 0_g \\ 4_u \\ 1_h \end{pmatrix}
\end{array}
$$

$$
\begin{array}{ccc}
|p.(0)\rangle & |p.(1)\rangle & |p.(2)\rangle \\
\begin{pmatrix} 10_{(ij)} \\ 10_{(ij)} \\ 0_d \\ 0_s \\ 0_{(gu)} \\ 0_{(gu)} \\ 0 \end{pmatrix} &
\begin{pmatrix} 0_{(ij)} \\ 0_{(ij)} \\ 1_d \\ 2_s \\ 0_{(gu)} \\ 0_{(gu)} \\ 0 \end{pmatrix} &
\begin{pmatrix} 0_{(ij)} \\ 0_{(ij)} \\ 0_d \\ 0_s \\ 1_{(gu)} \\ 1_{(gu)} \\ 0 \end{pmatrix}
\end{array}
$$

The full dynamics of the given CAO$_I$ is determined by no more than three CST steps.

As previously mentioned, the values of the state vector components determine only the *meaning* of the CAO as a multicomponent formation. The semantic interpretation of the transformations is also determined by the position of each CÆ in the CAO structure as well as by the type and parameters of the cardinal semantic operators and their semantic connectivity (topology).

### 5. Conclusion

The paper proposes, for the first time, a representation of an arbitrary semantic numeration system as a linear discrete dynamic system with nonlinear control based on a state vector.

A representation of a number/multinumber has been developed as a process of unfolding of meanings on the structure of a numeration method. The acquired *sense* of a multinumber emerges from the unity of its *meanings* and *structure*. Without taking into account the structure and parameters of the numeration method, the semantic unity disintegrates, degenerating into a mere set of meanings (a multiset).

The work does not address the cyclic topology of STop, the topology with feedback, possible cardinal additions to CÆs in the CST process, and the inclusion/exclusion of CÆs in the CAO, i.e. changes in the dimensionality of the dynamic system.

Further research in these areas remains important for future works.